\documentclass[12pt]{article}
\def\ben{\begin{equation}}
\def\een{\end{equation}}

\def\nn{\nonumber} \def\bd{\begin{document}} \def\ed{\end{document}}
\def\ds{\documentstyle} \let\fr=\frac \let\bl=\bigl \let\br=\bigr
\let\Br=\Bigr \let\Bl=\Bigl
\let\bm=\bibitem
\let\na=\nabla
\let\pa=\partial \let\ov=\overline
\newcommand{\be}{\begin{equation}}
\newcommand{\ee}{\end{equation}}
\def\ba{\begin{array}}
\def\ea{\end{array}}
\def\ft#1#2{{\textstyle{\frac{\scriptstyle #1}{\scriptstyle #2} } }}
\def\fft#1#2{{\frac{#1}{#2}}}
\def\del{\partial}
\def\vp{\varphi}
\def\sst#1{{\scriptscriptstyle #1}}
\def\oneone{\rlap 1\mkern4mu{\rm l}}
\def\td{\tilde}
\def\wtd{\widetilde}
\def\ie{{\it i.e.\ }}
\def\dalemb#1#2{{\vbox{\hrule height .#2pt
        \hbox{\vrule width.#2pt height#1pt \kern#1pt
                \vrule width.#2pt}
        \hrule height.#2pt}}}
\def\square{\mathord{\dalemb{6.8}{7}\hbox{\hskip1pt}}}
\newcommand{\ho}[1]{$\, ^{#1}$}
\newcommand{\hoch}[1]{$\, ^{#1}$}
\newcommand{\bea}{\setlength\arraycolsep{2pt} \begin{eqnarray}}
\newcommand{\eea}{\end{eqnarray}}
\newcommand{\ra}{\rightarrow}
\newcommand{\lra}{\longrightarrow}
\newcommand{\Lra}{\Leftrightarrow}
\newcommand{\bp}{\tilde \beta^\prime}
\newcommand{\tr}{{\rm tr} }
\newcommand{\Tr}{{\rm Tr} }
\def\0{{\sst{(0)}}}
\def\1{{\sst{(1)}}}
\def\2{{\sst{(2)}}}
\def\3{{\sst{(3)}}}
\def\4{{\sst{(4)}}}
\def\5{{\sst{(5)}}}
\def\6{{\sst{(6)}}}
\def\7{{\sst{(7)}}}
\def\8{{\sst{(8)}}}
\def\m{{\sst{(m)}}}
\def\n{{\sst{(n)}}}
\def\cA{{{\cal A}}}
\def\cB{{{\cal B}}}
\def\cF{{{\cal F}}}
\def\cG{{{\cal G}}}
\def\cH{{{\cal H}}}
\def\tV{\widetilde V}
\def\tW{\widetilde W}
\def\tH{\widetilde H}
\def\tE{\widetilde E}
\def\tF{\widetilde F}
\def\tA{\widetilde A}
\def\im{{{\rm i}}}
\def\tY{{{\wtd Y}}}
\def\ep{{\epsilon}}
\def\vep{{\varepsilon}}
\def\bD{{{\bar D}}}
\def\R{{{\mathbb R}}}
\def\C{{{\mathbb C}}}
\def\H{{{\mathbb H}}}
\def\CP{{{\mathbb C}{\mathbb P}}}
\def\RP{{{\mathbb R}{\mathbb P}}}
\def\Z{{{\mathbb Z}}}
\def\bA{{{\mathbb A}}}
\def\bB{{{\mathbb B}}}
\def\bC{{{\mathbb C}}}
\def\bD{{{\mathbb D}}}
\def\bE{{{\mathbb E}}}
\def\bZ{{{\mathbb Z}}}
\def\Re{{{\frak{Re}}}}
\def\Im{{{\frak{Im}}}}
\def\cosec{{\,\hbox{cosec}\,}}
\def\Gm{{\Gamma_{\!\! -}}}
\def\Gp{{\Gamma_{\!\! +}}}
\def\stan{{standard }}
\def\nonstan{{supernumerary }}
\def\p{{\partial}}
\def\kdel#1{{\fft{\del}{\del#1}}}
\def\bog{{Bogomolny }}
\def\om{{\omega}}
\newcommand{\nnr}{\nonumber \\}
\newcommand{\pd}{\partial}
\newcommand{\ud}{\textrm{d}}
\newcommand{\dTH}{T^{\prime \, 0}_\textrm{H}}
\newcommand{\dOi}{\Omega^{\prime \, 0}_i}
\newcommand{\bx}{{\bf x}}
\begin{document}
\begin{titlepage}

\vfill

\begin{flushright}
KIAS-P09017
\end{flushright}

\vfill

\begin{center}
   \baselineskip=16pt
   {\Large\bf Dyonic solution of Ho\v{r}ava-Lifshitz Gravity}
   \vskip 2cm
      Eoin \'{O} Colg\'{a}in and Hossein Yavartanoo
         \vskip .6cm
      \begin{small}
      $^1$\textit{Korea Institute for Advanced Study, \\
        Seoul 130-722, Korea}
        \end{small}\\*[.6cm]
\end{center}

\vfill
\begin{center}
\textbf{Abstract}

\end{center}
\begin{quote}
Recently proposed Horava-Lifshitz gravity promises a UV completion of Einstein's theory by sacrificing general covariance at short distances and introducing anisotropic spacetime scaling. Here we present a dyonic solution by coupling this  theory to a vector field and we discuss some properties of solution.

\end{quote}

\vfill

\end{titlepage}

We briefly review the Ho\v{r}ava-Lifshitz gravity theory \cite{horava09}, which builds on some of Ho\v{r}ava's previous work, in particular \cite{c3}.  The dynamical variables are $N, N^{i}$, sometimes referred to as the ``lapse" and ``shift" parameters from general relativity, and a $D$-dimensional spatial metric $g_{ij}$.  The theory allows anisotropic scaling with dynamical exponent $z$, 
\be
\mathbf{x} \rightarrow b \mathbf{x}, \quad t \rightarrow b^{z} t,
\ee
with Lorentz invariance reinstated for $z=1$. For general $z$, the classical scaling dimensions of the fields are 
\be
[g_{ij}] = 0, \quad [N_{i}] = z-1, \quad [N]=0. 
\ee
With its anisotropic scaling, the theory admits a foliation of D+1 spacetime, with the leaves of the foliation being hypersurfaces at constant time $t$. In general for $z=D$, the theory is expected to give a ghost-free UV-renormalisable theory of non-relativistic gravity in flat space. 

The metric in ADM decomposition \cite{adm} may be given by
\be
ds^2= - N^2  dt^2 + g_{ij} (dx^i - N^i dt) (dx^j - N^j dt). 
\ee
The action of the theory $S = S_{K} + S_{P}$ may be separated into kinetic terms 
\be
S_K = \frac{2}{\kappa^2} \int  dtd^3  \sqrt{g}N (K_{ij}K^{ij} -\lambda
K^2),
\ee
where the extrinsic curvature $K_{ij}$ is 
\be
K_{ij} = \fft{1}{2N} (\dot g_{ij} - \nabla_i N_j - \nabla_j N_i), 
\ee 
and potential terms $S_{P}$.
 
Completing the action involves adding potential terms which are of dimension equal to or less than the dimension of the kinetic term, $[K_{ij} K^{ij}] = 2z$. It is the presence of these relevant operators, added through the potential term, that govern how Lorentz invariance is restored in the IR. Ho\v{r}ava also subjected the theory to the extra requirement of ``detailed balance" \cite{horava09} as a means to whittle down the choice of these numerous relavant operators. 

The appropriate selection of these relevant operators so as to correctly recover general relativity is an issue that requires further study. For some recent musings on recovering general relativity at different scales by relaxing the detailed balance condition, see \cite{Nastase:2009nk}. For other works in this nascent area, including the implications for cosmology, see \cite{otherworks,cosmo}.      

For the moment, we continue the review of Ho\v{r}ava's original incarnation of the theory and persist with the detailed balance condition, which we will relax again later. Under this assumption, and specialising to $z=D=3$, the complete action of Ho\v{r}ava-Lifshitz is given by \cite{horava09}
\bea
S_{hl} &=&\int dtd^3\bx({\cal L}_0 + {\cal L}_1),\nn\\ {\cal L}_0
&=& \sqrt{g}N\left\{\frac{2}{\kappa^2}(K_{ij}K^{ij} -\lambda
K^2)+\frac{\kappa^2\mu^2(\Lambda_{W}  R
  -3\Lambda_{W} ^2)}{8(1-3\lambda)}\right\}\,,\nn\\ {\cal L}_1&=&
\sqrt{g}N\left\{\frac{\kappa^2\mu^2 (1-4\lambda)}{32(1-3\lambda)}R^2
-\frac{\kappa^2}{2w^4} \left(C_{ij} -\frac{\mu w^2}{2}R_{ij}\right)
\left(C^{ij} -\frac{\mu w^2}{2}R^{ij}\right)\right\},\label{action}
\eea 
where $\lambda$ and $\kappa$ are dimensionless constants and the Cotton tensor - a measure of conformal flatness in $D=3$ - is defined by
\be
C^{ij}=\epsilon^{ik\ell}\nabla_k\left(R^j{}_\ell
-\frac14R\delta_\ell^j\right).
\ee
All spatial indices may be raised and lowered using $g_{ij}$. The equations of motion were independently worked out in \cite{Lu:2009em} and \cite{c2}, but owing to their length, we omit them. 

One special feature of this theory are emerging constants. Comparing ${\cal L}_0$ to that of general relativity in the ADM
formalism\footnote{The Einstein-Hilbert action in ADM formalism is given by \[
S_{EH} = \fft{1}{16\pi G} \int d^4x \sqrt{g} N (K_{ij} K^{ij} - K^2 + R -
2\Lambda). \]}, the speed of light, Newton's constant and the cosmological
constant are:
\be
c=\fft{\kappa^2\mu}{4} \sqrt{\fft{\Lambda_W }{1-3\lambda}}\,,\qquad
G=\fft{\kappa^2}{32\pi\,c}\,,\qquad
\Lambda=\ft32 \Lambda_w.\label{cg}
\ee
Strictly speaking, to make the comparison between ${\cal L}_{0}$, and Einstein-Hilbert action, we must take $\lambda = 1$, however in Ho\v{r}ava-Lifshitz gravity $\lambda$ represents a dynamical coupling constant. Adopting the range $\lambda > 1/3$, one sees that the cosmological constant $\Lambda$ is necessarily negative. Despite this, one may analytically continue $\mu \rightarrow i \mu, w^2 \rightarrow -i w^2$, to make the cosmological constant positive.

We consider this system coupled to a vector field. A general non-relativistic action $S_{m}$ where the potential depends only on gauge field and field strength was presented in \cite{c2}
\bea
S_{m} &=& \int  dtd^3\bx {\cal L}_{2}, \nn \\
&=& -{1\over 4g^2}\int
d^3xdt \sqrt{g}N\bigg[-{2\over
N^2}g^{ij}(F_{0i}-N^kF_{ki})(F_{0j}-N^lF_{lj}) \cr\cr
&-& {m^2\over N^2}(A_0-N^iA_i)(A_0-N^jA_j)
+G[F_{ij}F^{ij}, A_iA^i]\bigg].
\eea
Here $F_{0i}=\partial_t A_i-\partial_iA_0$, $F_{ij}= \partial_i A_j-\partial_jA_i$ is the field strength.
The scaling dimensions of the fields are $[A_i] =0, [A_0] = 2$. 
For renormalisability in the UV we may consider the  function $G$  to be, at most, cubic in $F^2$, but here, for simplicity we ignore higher derivative terms, considering solely $G = F_{ij} F^{ij}$ and also $m=0$. 

We are interested in static, spherically symmetric solutions and adopt the same metric ansatz that appeared in \cite{Lu:2009em,Cai:2009pe}, 
\bea 
ds^2 &=& -N(r)^2dt^2 + \frac{dr^2}{f(r)} + r^2 (d\theta^2+\sin^2\theta d\phi^2),  \nnr
F_{r0} &=& A(r)', \quad F_{\theta \phi} = p \sin \theta, 
\eea 
where the flux ansatz is chosen to respect the $SO(3)$ action on the $S^2$ and to satisfy the equations coming from varying $S_m$ with respect to $A_{0}$ and $A_i$ respectively:
\bea
\label{fields}
\frac{1}{\sqrt{g}}\left( \frac{ \sqrt{g}  f  A' }{N } \right)' &=& 0, \nnr
\partial_{\theta} (\sqrt{g}   F^{\theta \phi}) &=& \partial_{\phi} (\sqrt{g}  F^{\theta \phi}) = 0. 
\eea  
For this choice of ansatz, the Cotton tensor disappears $C_{ij}=0$ and the vanishing of the $N_{i}$ also present a considerable simplification. Considering only the Lagrangian ${ \cal L}_{0}$, one obtains the AdS Schwarzchild black hole solution
\be
\label{schads}
N^2 = f = 1-\frac{\Lambda_{W}}{2} r^2 - \frac{M}{r}. 
\ee

The general solution may be most easily obtained by by-passing the equations of motion and instead placing the ansatz into the full Lagrangian ${\cal L}_{0} + {\cal L}_{1} + {\cal L}_{2}$.  We focus on the $\lambda=1$ limit of the Ho\v{r}ava-Lifshitz action where usual Einstein description of gravity should be restored. As may be seen from (\ref{fields}), when $m=0$,  one has 
\be
A(r)' = \frac{N(r) q}{\sqrt{f(r)} r^2}
\ee
with $q$ a constant introduced, which we will later confirm to be the electric charge. Substituting this expression into the Lagrangian, the equations of motion may be determined from varying the following reduced action with respect to $N$ and $f$. Up to an overall factor, the reduced one-dimensional action is 
\bea
{\cal L} &=& - \frac{\kappa^2 \mu^2}{4(3 \lambda-1)}\fft{N}{\sqrt{f}}\Biggl(2 - 3\Lambda_W r^2 -2 f - 2r f' +
\fft{\lambda-1}{2\Lambda_W} f'^2 - \fft{2\lambda (f-1)}{\Lambda_W r}f' \nnr
&+& \fft{(2\lambda-1)(f-1)^2}{\Lambda_W r^2}\Biggr) + \frac{1}{g^2} \fft{2 N}{r^2 \sqrt{f}} (q^2+p^2).
\eea
The resulting equations of motion are satisfied for the following solution
\bea
\label{m0sol}
f &=& 1-\Lambda_{W} r^2 - \sqrt{ \frac{8( q^2 + p^2)}{(\kappa \mu g)^2} + \alpha^2 r}, \nnr
N &=& \sqrt{f}, \quad A = -\frac{q}{r},  
\eea 
where $\alpha^2$ is an integration constant, which is up to an additive constant, the mass \cite{Cai:2009pe,Banados:1993ur}. Note that when $p=q=0$, we recover the solution of \cite{Lu:2009em}, as expected. Here one can can confirm both $q$ and $p$ as the electric and magentic charge respectively by preforming the following integrals over the two-sphere,
\be
q = \frac{1}{4 \pi} \int_{S^2} F vol(S^2), \quad p = \frac{1}{4 \pi} \int_{S^2} \star F vol(S^2). 
\ee

We note that the solution is asymptotically $AdS_4$, but there is a horizon at the largest root of f(r). 
The Hawking temperature of the black hole is given by
\bea
T &=& \frac{1}{4 \pi} f(r)' |_{r=r_{+}}, \nnr
&=& \frac{-3 \Lambda_w r_+^2 -1}{8 \pi r_+} + \frac{1}{(\kappa \mu g)^2} \frac{(q^2+p^2)}{\pi r_+ (1-\Lambda_w r_+^2)}, 
\eea 
where $r_+$ the largest root of $f$ determines the location of the horizon. 

Bearing in mind that $\Lambda < 0$, we note that there is an extremal limit with $T=0$ when 
\be
-\Lambda_w r_{+}^2 = -\frac{1}{3} + \frac{2}{3} \sqrt{1-\frac{6(p^2+q^2)}{(\kappa \mu g)^2}}. 
\ee

Following \cite{Lu:2009em}, it is possible to relax this detailed balance condition by considering the Lagrangian 
\be
{\cal L} = {\cal L}_{0} + (1-\epsilon^2) {\cal L}_{1},
\ee
where $\epsilon$ represents a slight deviation. With this slight adjustment, one may repeat the previous analysis and find the solution 
\bea 
f &=& 1- \frac{\Lambda r^2}{1-\epsilon^2} - \frac{1}{(1-\epsilon^2)} \sqrt{[ \frac{8(q^2+p^2)}{(\kappa \mu g)^2} + \alpha^2r](1-\epsilon^2) + \epsilon^2 \Lambda^2 r^4  }, \nnr
N^2 &=& f, \quad A = - \frac{q}{r}. 
\eea

The large distance behaviour of the function is given by 
\be
f = 1 - \frac{\Lambda r^2}{1+\epsilon}+\frac{\alpha^2}{2 \epsilon r \Lambda} + {\cal O}(\frac{1}{r^4}).  
\ee
Here, for non-vanishing $\epsilon$ i.e. away from detailed balance set-up, we see the metric has a finite mass by comparing with (\ref{schads}). This mass diverges for the detailed balance value $\epsilon = 0$, in which case we recover (\ref{m0sol}). In the other limit where $\epsilon = 1$, ${\cal L}_1$ disappears from the Lagrangian and one gets this solution in AdS space

\be
N^2 = f = 1-\frac{\Lambda_{W}}{2} r^2 - \frac{M}{r}.+ \frac{4(q^2+p^2)}{r \Lambda (\kappa \mu g)^2}.  
\ee

\end{document}